\documentclass[a4paper,11pt]{article}
\usepackage{pos}
\usepackage{amsmath,bm,color,fancyhdr,hyperref,lastpage,units,soul,caption}
\usepackage{graphicx}
\graphicspath{figures/}

\title{Charm fluctuations in (2+1)-flavor QCD at high temperature}

\author*[a,1]{Sipaz Sharma}

\affiliation[a]{Fakult\"at f\"ur Physik, Universit\"at Bielefeld, D-33615 Bielefeld, Germany}
\note{For the HotQCD Collaboration.}

\emailAdd{sipaz@physik.uni-bielefeld.de}

\abstract{Using the high statistics datasets of the HotQCD Collaboration,
	generated with the HISQ (2+1)-flavor action for light and strange quarks, 
	and treating the charm sector in the quenched approximation, we analyze 
	the second and fourth order cumulants of charm fluctuations and 
	the correlations of charm with lighter conserved flavor quantum numbers. 
	We can make use of a factor 100 larger statistics on ${N_\tau =8}$ lattices which
	never have been used in studies of charm fluctuations. 
	
	Analyzing correlations of charm fluctuations with baryon number and 
	electric charge fluctuations we can project onto charmed baryon and 
	meson correlations and compare results with quark model extended hadron 
	resonance gas model calculations. We aim at a precise determination of 
	the dissociation temperature of charmed hadrons and will probe the 
	sensitivity of the fluctuations observables to the presence of 
	multiple-charmed baryons.}

\FullConference{%
  The 39th International Symposium on Lattice Field Theory (Lattice2022),\\
  8-13 August, 2022 \\
  Bonn, Germany 
}

\begin{document}
\maketitle

\section{Introduction}
Charmed hadrons have been proposed as an important probe in understanding properties as well as signatures of the deconfined medium consisting of quark-gluon plasma  formed in the heavy-ion collision experiments ever since the seminal work of Matsui and Satz \cite{Matsui:1986dk}. 

It is now well established that the strong interaction matter at vanishing baryon chemical potential undergoes restoration of the spontaneously broken chiral symmetry via a crossover transition since the small yet non-vanishing up and down quark masses result also in the explicit breaking of the ${SU(2)_L\times SU(2)_R}$ chiral symmetry group. This chiral crossover transition occurs at a pseudo-critical temperature, ${T_{pc}}$ which calculated using Lattice QCD formalism is  ${T_{pc}=156.5\pm1.5}$ MeV \cite{HotQCD:2018pds}. Interestingly, as per observations made in the heavy-ion collision experiments, deconfinement of light-quark hadrons also occurs at ${T_{pc}}$, and thus the relevant degrees of freedom change from partonic to hadronic in going from high temperature phase to temperatures below ${T_{pc}}$.
 
One of the open questions which heavy-ion experiments aim to answer is whether open-charm states melt beyond ${T_{pc}}$, or at ${T_{pc}}$ like other light-quark hadrons. Statistical Hadronization Model (SHM) is one of theoretical tools to model experimentally measured hadron abundances, and has been extended to SHMc: which incorporates multiple-charmed states based on the observation that the production of charmed-states is strongly Boltzmann suppressed given the massive charm quark mass in relation to the relevant temperature scale, and is instead a result of the initial hard collisions \cite{Andronic:2021erx}. Another approach to this problem is using lattice calculations of hadronic correlation functions \cite{Datta:2003ww},\cite{Ding:2012sp} to extract details of the charm hadron spectrum; or by comparing various charm cumulants calculated on the lattice with the Hadron Resonance Gas (HRG) model  predictions \cite{Allton:2005gk} -- which are valid only in the low temperature phase -- and hence considering the deviation of lattice results from the HRG calculations a signal of the open-charm sector melting \cite{BAZAVOV2014210}. Moreover, existence of the not-yet-discovered open-charm states can also be predicted by the comparison of lattice results with the HRG calculations \cite{BAZAVOV2014210}; and signals of the exotic charm states such as tetraquarks can shed light on the arrangement of quarks inside the bound states \cite{Bicudo:2022cqi}.
\section{Hadron Resonance Gas Model}
\subsection{Partial Pressure from the Open-Charm Sector}
HRG describes a non-interacting gas of hadron resonances, and therefore is valid in the hadronic phase below ${T_{pc}}$. It has been successful in describing the particle abundance ratios measured in the heavy-ion experiments. The grand-canonical partition function in the open-charm sector -- provided by the Hadron resonance gas model -- given by ${\mathcal{Z}^{HRG}_{C}(T,V,\overrightarrow{\mu})}$, is dependent upon temperature ${T}$, volume ${V}$, and chemical potentials ${\overrightarrow{\mu}=(\mu_B, \mu_Q, \mu_S, \mu_C)}$ -- where subscripts of the components of $\overrightarrow{\mu}$ correspond to various conserved quantum numbers such as baryon number ${B}$, electric charge ${Q}$, strangeness ${S}$ and charm ${C}$ respectively. Partial pressure ${P_C}$ from the open-charm sector is related to ${\text{ln}\mathcal{Z}^{HRG}_C}$ as follows,
\begin{equation}
	{\dfrac{P_C(T,V,\overrightarrow{\mu})}{T^4}=\dfrac{\text{ln}\mathcal{Z}^{HRG}_{C}(T,V,\overrightarrow{\mu})}{VT^3}}\text{ .}
	\label{eq:P-Z}
\end{equation}

The left-hand-side of Eq.\ref{eq:P-Z} can be further decomposed into separate contributions to partial open-charm pressure coming from the charmed-meson (C-mesons) and the charmed-baryon (C-baryons) sectors, denoted by ${M_C}$ and ${B_C}$ respectively:
\begin{equation}
{P_C(T,\overrightarrow{\mu})/T^4=M_C(T,\overrightarrow{\mu})+B_C(T,\overrightarrow{\mu})} \text{ .}
\end{equation}
Based on the calculations presented in References \cite{Allton:2005gk}, \cite{BAZAVOV2014210}, ${M_C}$ and ${B_C}$ take following forms,
\begin{gather}
\begin{aligned}
	{M_C(T,\overrightarrow{\mu})}&{=\dfrac{1}{2\pi^2}\sum_{i\in \text{C-mesons}}g_i \bigg(\dfrac{m_i}{T}\bigg)^2K_2(m_i/T)\text{cosh}(Q_i\hat{\mu}_Q+S_i\hat{\mu}_S+C_i\hat{\mu}_C)} \text{ ,}\\
	{B_C(T,\overrightarrow{\mu})}&={\dfrac{1}{2\pi^2}\sum_{i\in \text{C-baryons}}g_i \bigg(\dfrac{m_i}{T}\bigg)^2K_2(m_i/T)\text{cosh}(B_i\hat{\mu}_B+Q_i\hat{\mu}_Q+S_i\hat{\mu}_S+C_i\hat{\mu}_C)} \text{ .}
	\label{eq:McBc}
\end{aligned}
\end{gather}
In the above equations, at a given temperature ${T}$, summation is over all C-mesons/baryons with masses given by ${m_i}$; degeneracy factors of the states with equal mass and same quantum numbers are represented by ${g_i}$; to work with a dimensionless notation, chemical potentials are normalised by the temperature: ${\hat{\mu}_X = \mu/T}$, ${\forall X \in \{B, Q, S, C\}}$; ${K_2(x)}$ is a modified Bessel function, which for a large argument can be approximated by
${K_2(x)}\sim\sqrt{\pi/2x}\;e^{-x}\;[1+\mathbb{O}(x^{-1})]$\cite{Allton:2005gk}: consequently, if a charmed-state under consideration is much heavier than the relevant temperature scale, such that ${m_i\gg T}$, then the contribution to ${P_C}$ from that particular state will be exponentially suppressed, e.g., singly-charmed
 ${\Lambda}_c^{+}$ baryon has a PDG (Particle Data Group) mass of about $2286$ MeV, whereas doubly-charmed ${\Xi_{cc}^{++}}$ baryon's mass as tabulated in PDG records is about $3621$ MeV, therefore at ${T_{pc}}$, contribution to ${B_C}$ from ${\Xi_{cc}^{++}}$ will be suppressed by a factor of $10^{-3}$ in relation to ${\Lambda}_c^{+}$ contribution. In other words, in order to distinguish signals of multiple-charm states from ${C=1}$ sector, a precision study based on orders of magnitude larger statistics than that has been used in the present study would be required.
\subsection{Generalized Susceptibilities of the Conserved Charges}
\label{subsec: susc}
To project out the relevant degrees of freedom in the charm sector, one calculates the generalized susceptibilities, ${\chi^{BQSC}_{klmn}}$, of the conserved charges by taking appropriate derivatives of the total pressure ${P}$ -- which contains contribution from ${P_C}$ -- with respect to the chemical potentials of the quantum number combinations one is interested in:
\begin{equation} 
{\chi^{BQSC}_{klmn}=\dfrac{\partial^{(k+l+m+n)}\;\;[P\;(\hat{\mu}_B,\hat{\mu}_Q,\hat{\mu}_S,\hat{\mu}_C)\;/T^4]}{\partial\hat{\mu}^{k}_B\;\;\partial\hat{\mu}^{l}_Q\;\;\partial\hat{\mu}^{m}_S\;\;\partial\hat{\mu}^{n}_C}}\bigg|_{\overrightarrow{\mu}=0}\text{.}
\label{eq:chi}
\end{equation}
 In Equation \ref{eq:chi}, as long as ${n \neq 0}$ and ${(k+l+m+n) \in \text{even}}$, ${P}$ can be replaced by ${P_C}$, since the derivative w.r.t. ${\hat{\mu}_C}$ will always project onto the open-charm sector. In the following, if the subscript corresponding to a conserved charge is zero in the L.H.S. of Eq. \ref{eq:chi}, then both the corresponding superscript as well the zero subscript will be suppressed.
 
 Based on the following observations, it is possible to construct some useful combinations of the various generalized susceptibilities which can indirectly shed light on the properties of the open-charm sector:\\
 $\bullet$ Baryon-charm correlations,  ${\chi^{BC}_{mn}}$, will receive contributions from the singly, doubly and triply-charmed states carrying baryon number denoted by ${B_{C,1}}$,  ${B_{C,2}}$ and  ${B_{C,3}}$ respectively, but as argued earlier, dominant contribution to the open-charm partial pressure will come only from the $C=1$ sector, therefore,
 	\begin{equation}
 	 {\chi^{BC}_{mn}=B_{C,1}+2^nB_{C,2}+3^nB_{C,3}\simeq B_{C,1}}\text{.}
 	 \end{equation}
  $\bullet$
 Since baryons carry ${B=1}$, ratios such as ${\chi^{BC}_{mn}/\chi^{BC}_{kl}}$, will always be unity in the HRG phase irrespective of the details of the baryon mass spectrum, ${\forall\{(m+n), (k+l)\} \in \text{even}}$.\\
 $\bullet$
 Ratios like ${\chi^{BC}_{1n}/\chi^{BC}_{1l}}$ should be unity throughout the entire temperature range, ${\forall \{n,l\} \in \text{odd}}$.
 \section{Simulation Details}
 \subsection{Lattice Setup}
 We have used approximately one-third of the available (2+1)-flavor HISQ (Highly Improved Staggered Quark) configurations generated by the HotQCD collaboration -- tabulated in Tab. I of Ref. \cite{Bollweg:2021vqf} -- for the physical strange to light quark mass ratio, ${m_s/m_{l}}=27$, at a temporal lattice extent, ${N_{\tau}}=8$. To set the temperature scale, $f_K$ scale setting from Ref. \cite{Bollweg:2021vqf} has been used. Charm-quark sector has been treated in the quenched approximation. Details of the charm mass tuning and its parametrization will be discussed in the next subsection \ref{sec:mass_tunning}.
 
 The partition function of QCD with two light, one strange and one charm quark flavors with ${m_l}$, ${m_s}$ and ${m_c}$ being their respective masses is given by:
 \begin{equation}
 	{\mathcal{Z}=\displaystyle\int \mathcal{D}[U] \{\text{det } D(m_{l})\}^{2/4}\{\text{det } D(m_{s})\}^{1/4}\{\text{det } D(m_{c})\}^{1/4}  e^{-S_{g}}}\text{.}
 	\label{eq: Z}
 \end{equation} 
Here, ${U}$ are the ${SU(3)}$ gauge links; ${S_g}$ is the gluonic part of the action; ${D(m_i)}$ for ${i\in(l,s,c)}$ represent the fermionic matrices; determinants of the fermionic matrices are raised to powers of $1/4$ because we are dealing with the staggered quarks. As pointed out in Eq. \ref{eq:P-Z}, the partition function of Eq. \ref{eq: Z} can be related to the total pressure, which can be further used to calculate the generalized susceptibilities in the $BQSC$ basis, presented in Eq. \ref{eq:chi}. Calculation of $\chi^{BQSC}_{klmn}$ involves derivatives of the pressure, and on the lattice this is achieved by the stochastic estimation of various traces -- consisting of inversions and derivatives  of the fermionic matrices -- using random noise method. In particular, $500$ random vectors have been used to calculate various traces per configuration, except for Tr ${\big(D^{-1}\frac{\partial D}{\partial \mu}\big)}$ -- which turned out be particularly noisy and therefore, $2000$ random vectors have gone into its calculation. At present, we have gone upto fourth order in calculating various charm cumulants.

 \begin{figure}[!h]
 	\includegraphics[width=0.52\textwidth]{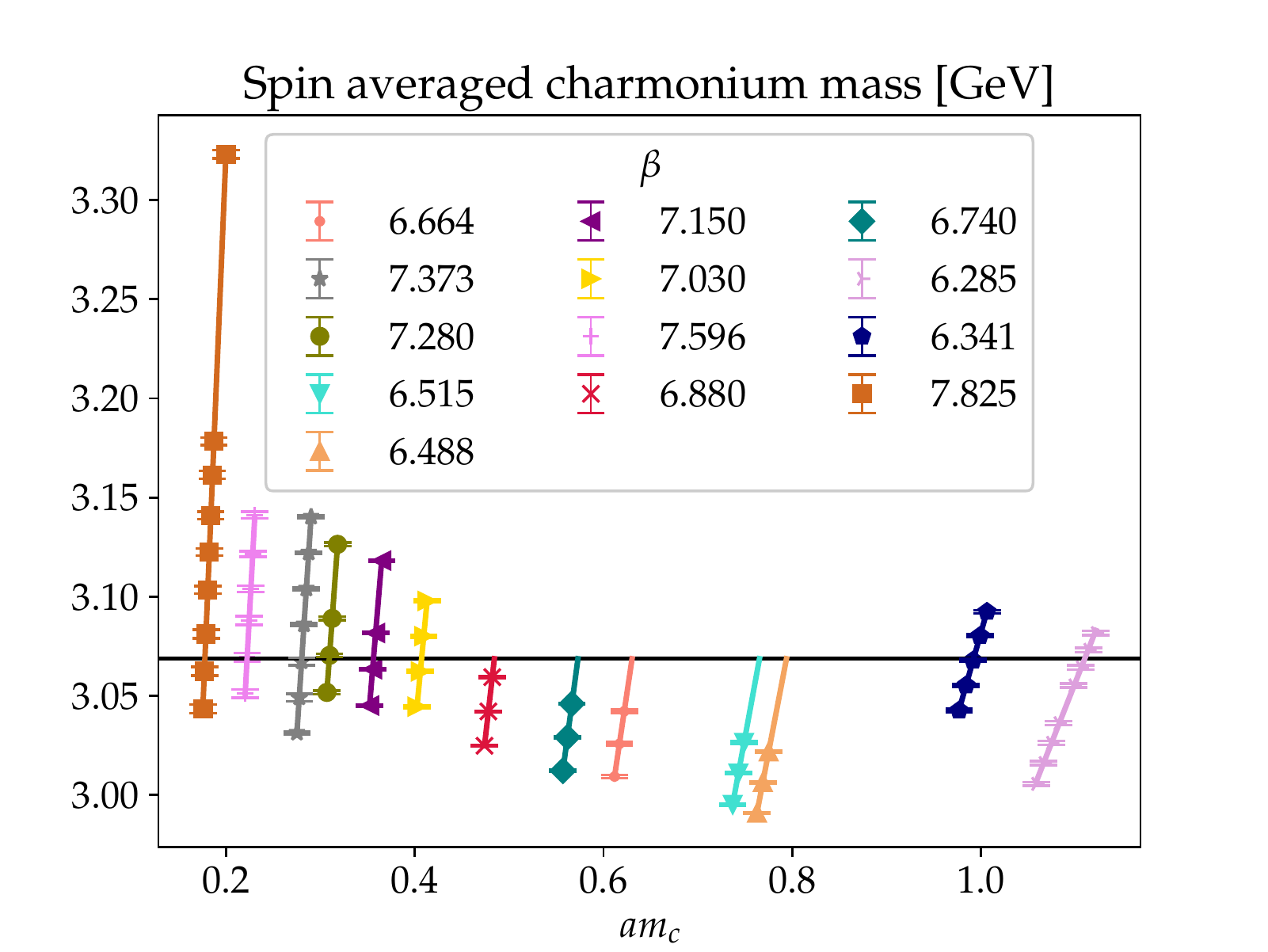}%
 	\hspace{0.01\textwidth}%
 	\includegraphics[width=0.52\textwidth]{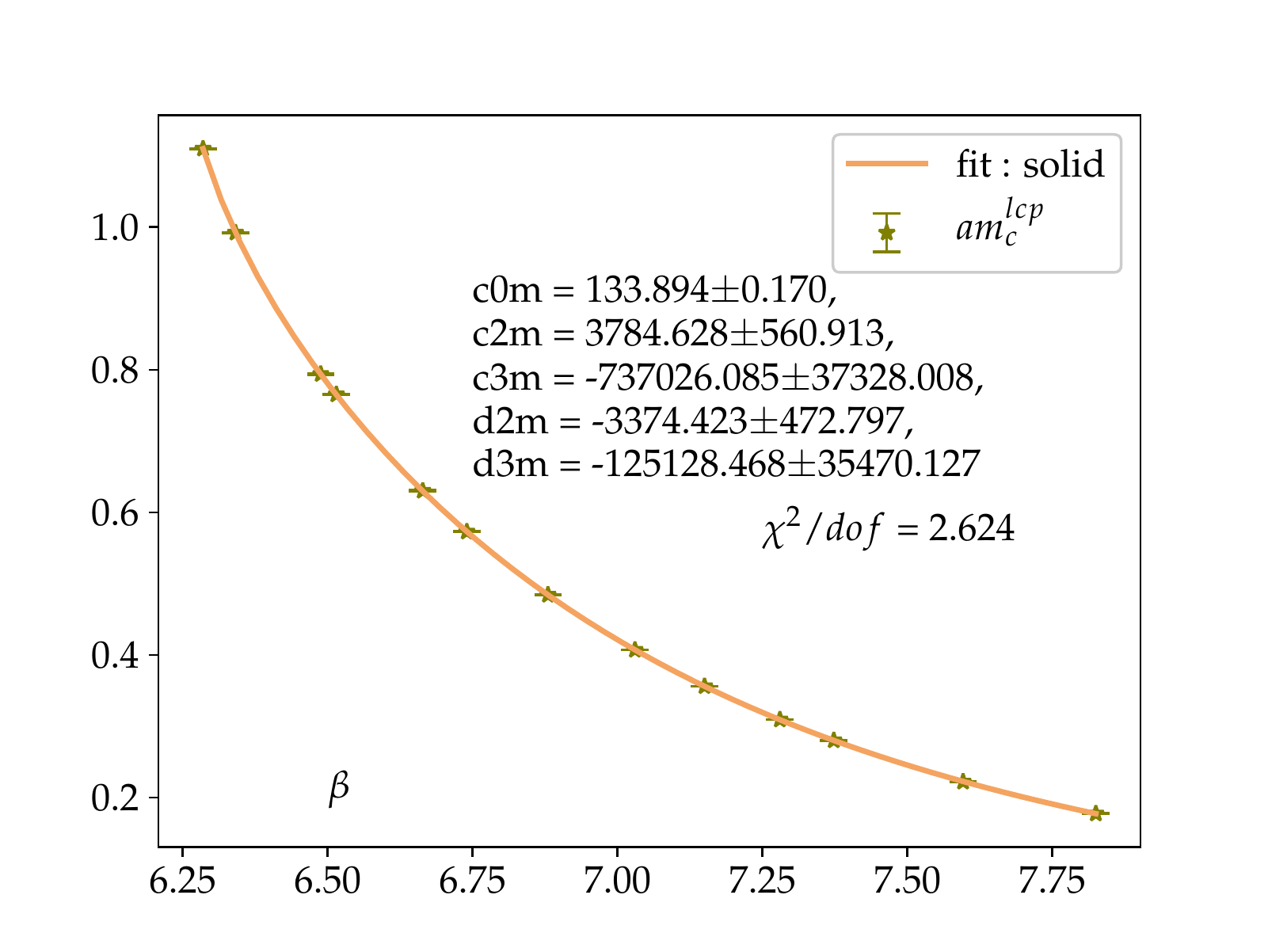}%
 	\caption{Charm mass tuning -- by considering the PDG spin-averaged charmonium mass, $(3m_{J/\psi}+m_\eta)/4$ as the LCP -- for thirteen $\beta$ values using various trial $am_c$ values is shown [left]. RG inspired fit using Eq. \ref{eq:RG} to the intersections of LCP with the linear interpolations of each beta value, $am_{c}^{lcp}(\beta)$ is shown [right]; various fit coefficients are also quoted in the figure.}
 	\label{fig:mass_tuning}
 \end{figure}
 
 \subsection{Charm Mass Tuning on the Lattice}
 \label{sec:mass_tunning}
 For each of the thirteen beta values, $\beta \in [6.285,7.825]$, more than three trial values of the bare charm quark mass, $am_c$ have gone into the calculation of the spin-averaged charmonium masses given by ${(3m_{J/\psi}+m_{\eta_{c\bar{c}}})/4}$, which are represented by different colors as well markers for each beta value in Fig. \ref{fig:mass_tuning} [left]. $f_K$ scale setting from Ref. \cite{Bollweg:2021vqf} has been used to convert various spin-averaged charmonium masses in lattice units to GeV. Linearly interpolated solid lines for each beta intersect the black-horizontal Line of Constant Physics (LCP) defined by the PDG ${(3m_{J/\psi}+m_{\eta_{c\bar{c}}})/4} = 3.06865 $ GeV at $am_{c}^{lcp}(\beta)$ in Fig. \ref{fig:mass_tuning} [left]. These intersections -- the olive-stars -- are then fitted to the following Renormalization Group (RG) inspired ansatz in Fig. \ref{fig:mass_tuning} [right], 
 
 \begin{equation}
 	{am_c^{lcp}(\beta)= \bigg(\dfrac{20b_0}{\beta}\bigg)^{\frac{4}{9}} c_{0m} f(\beta)\;\bigg[\;\dfrac{1+c_{2m}(\frac{10}{\beta})f^2(\beta)+c_{3m}(\frac{10}{\beta})f^3(\beta)}{1+d_{2m}(\frac{10}{\beta})f^2(\beta)+d_{3m}(\frac{10}{\beta})f^3(\beta)}\;\bigg]} \text{.}
 	\label{eq:RG}
 \end{equation}
Here, $f(\beta)$ is the 2-loop QCD beta function,
\begin{equation}
	f(\beta)=\bigg(\dfrac{10b_0}{\beta}\bigg)^{-b_1/2b_0^2}\text{ exp }(-\beta/(20b_0)) \text{,} \label{eq:2-loop}
\end{equation} 
with $b_0$ and $b_1$ being the perturbative expansion coefficients of the QCD $\beta$-function for number of colors, $N=3$ and number of flavors, $N_f=3$ ,
 \begin{gather}
 	\begin{aligned}
 		b_0&=\dfrac{1}{(4\pi)^2}\bigg(\dfrac{11}{3}N-\dfrac{2}{3}N_f\bigg)\text{ ,}\\
 		b_1&=\dfrac{1}{(4\pi)^4}\bigg(\dfrac{34}{3}N^2-\dfrac{10}{3}NN_f-\dfrac{N^2-1}{N}N_f\bigg) \text{ .} \label{C7-B}	
 	\end{aligned}
 \end{gather}
 Fig. \ref{fig:mass_tuning} [right] quotes the value of the fit coefficients in Eq. \ref{eq:RG} as well the $\chi^2/dof$ which indicates the quality of the fit.

 \section{Results and Conclusions}
 \begin{figure}[!h]
 	\includegraphics[width=0.52\textwidth]{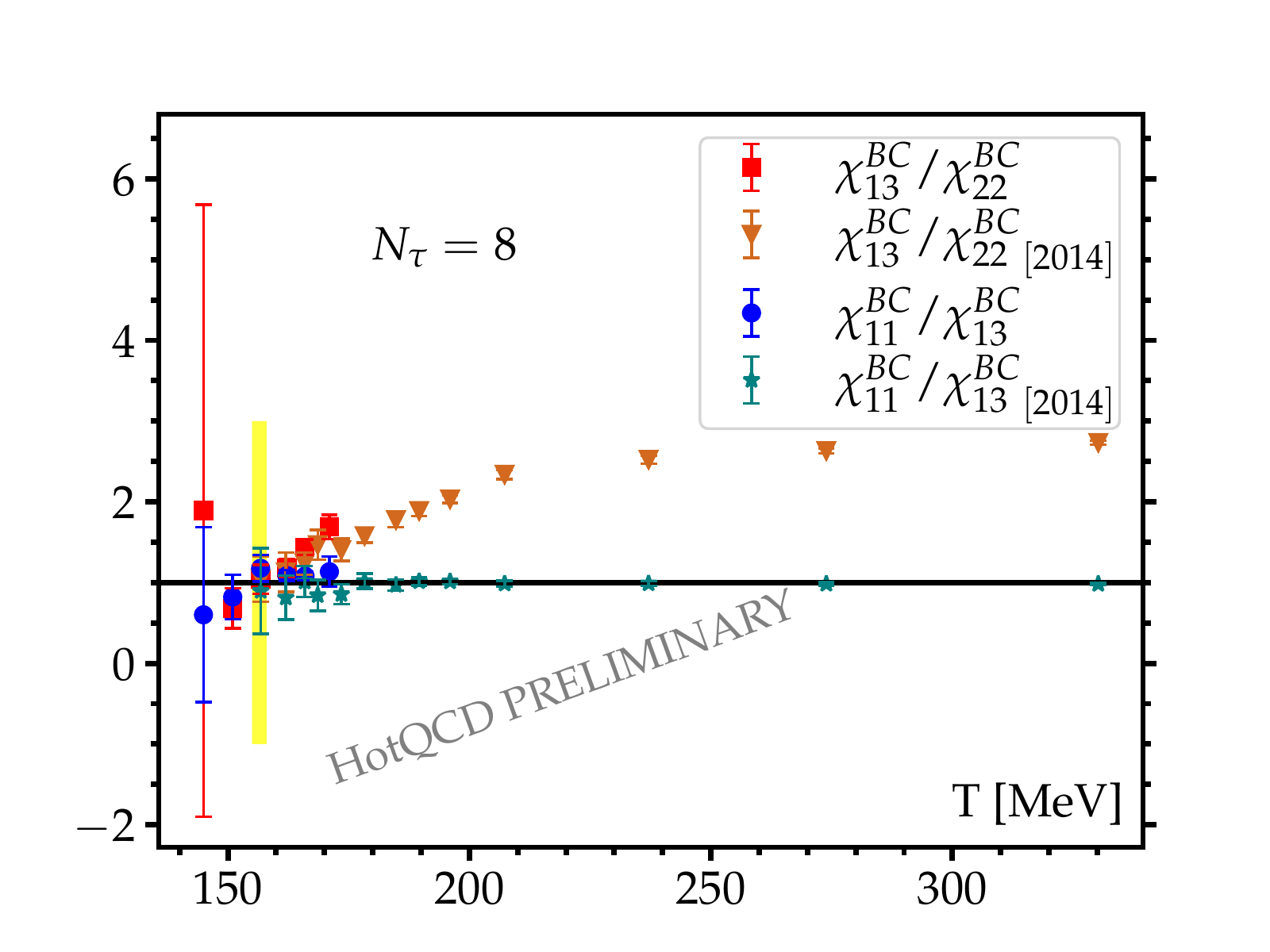}%
 	\hspace{0.01\textwidth}%
 	\includegraphics[width=0.52\textwidth]{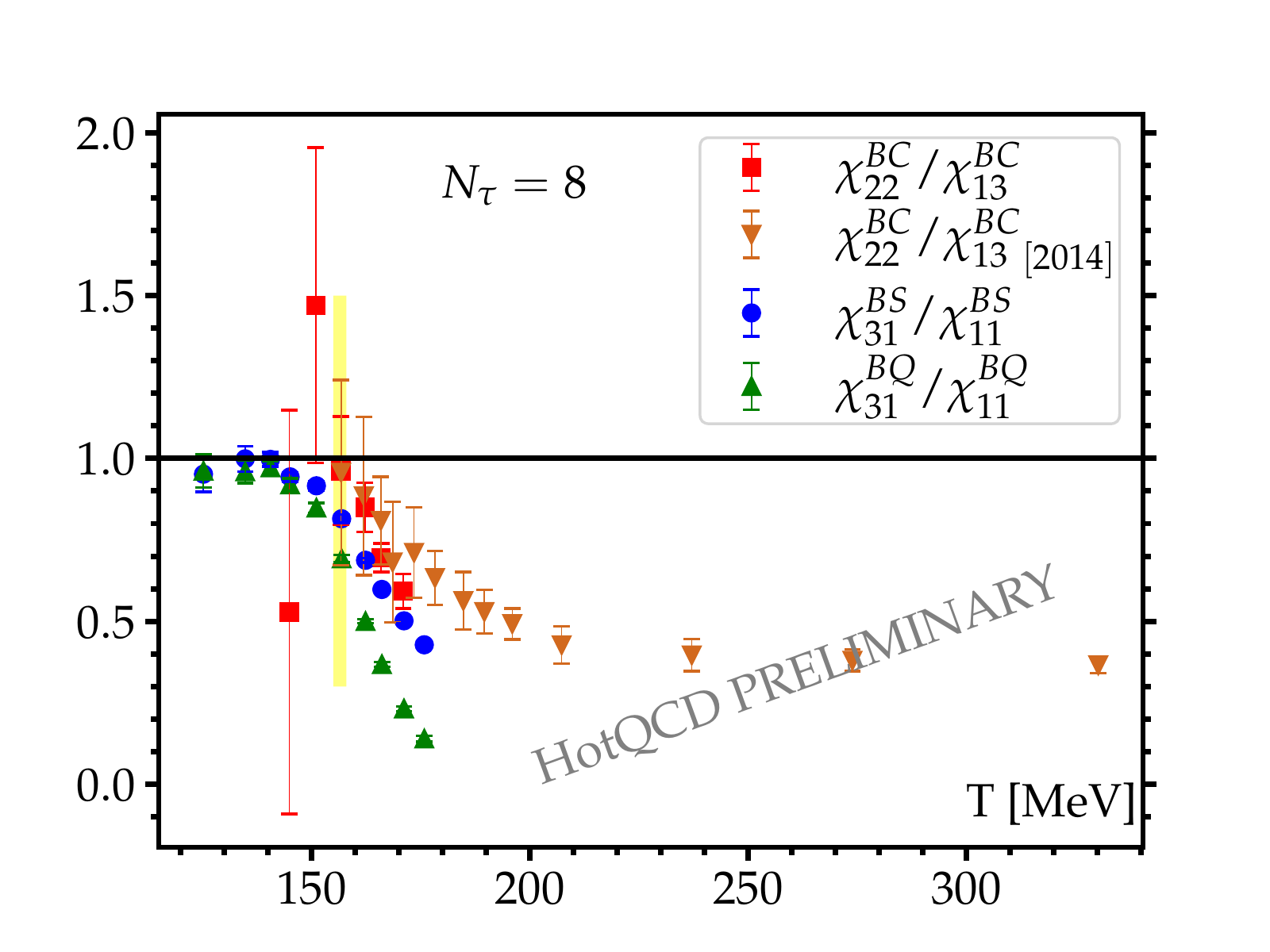}%
 	\caption{Ratios of various $BC$ correlations [left]. Ratios of $BC$, $BS$, $BQ$ correlations [right].}
 	\label{fig:BC-ratios}
 \end{figure}
As pointed out in Subsec.\ref{subsec: susc}, the ratios of the baryon-charm correlations, e.g., $\chi^{BC}_{13}/\chi^{BC}_{22}$ -- which comprise of different orders of  derivative w.r.t $\hat{\mu}_B$, such that the number of derivatives w.r.t. $\hat{\mu}_C$ are adjusted based on the availability of the highest order cumulant -- provide an indication about the onset of the melting of the open-charm baryon states by deviating from their low temperature HRG value -- which is unity. Fig \ref{fig:BC-ratios} [left] shows the updated $\chi^{BC}_{13}/\chi^{BC}_{22}$ values with the red-square markers for $T\in(144.95,171.02)$ MeV; in addition to that Fig \ref{fig:BC-ratios} [left] is augmented with the high temperature data, $T\in(156.8,330.2)$ MeV -- the brown-downward-triangles -- from the 2014 HotQCD analysis \cite{BAZAVOV2014210}, which took into account an order of magnitude smaller statistics at $T_{pc}$, and this fact reflects in the errors of the data points. Combination of the low-temperature and the high-temperature data sets, as well as the shrinkage of the yellow band representing $T_{pc}$ based on Reference \cite{HotQCD:2018pds}, enables discerning the deviation of $\chi^{BC}_{13}/\chi^{BC}_{22}$ from unity right after $T_{pc}$, which in turn hints at the change of the carriers of the charm degrees-of-freedom at $T_{pc}$. 

Given the massive charm quark, ratios of the baryon-charm correlations containing one derivative w.r.t. $\hat{\mu}_B$ should stay unity for the entire temperature range, and the blue-circles representing the updated $\chi^{BC}_{11}/\chi^{BC}_{13}$ values as well as the teal-stars depicting the 2014 HotQCD results in Fig \ref{fig:BC-ratios} [left] are a clear testament of this fact.

Fig. \ref{fig:BC-ratios} [right] depicts the reciprocal of $\chi^{BC}_{13}/\chi^{BC}_{22}$ for both the updated and the 2014 HotQCD data, therefore the above discussion applies. Ratios of the baryon-electric-charge as well as the baryon-strangeness correlations separated by even $\hat{\mu}_B$ derivatives should be unity in the HRG phase, but should show deviation from unity when the relevant degrees of freedom are no longer baryons. In Fig. \ref{fig:BC-ratios} [right], unlike the red-squares which take into account one-third of the available statistics, the green-triangles representing $\chi^{BQ}_{31}/\chi^{BQ}_{11}$ and the blue-circles representing $\chi^{BS}_{31}/\chi^{BS}_{11}$ with relatively smaller errors are calculated by taking into account all the available (2+1)-flavor HISQ configurations generated by the HotQCD collaboration. Although, the electrically-charged-baryon sector as well as the strange-baryon sector start deviating from the HRG expectation slightly before $T_{pc}$, the charmed-baryon sector shows clear deviation from unity right at $T_{pc}$. Notice that the green-triangles and the blue-circles do not incorporate the charmed-states, therefore it would be interesting to see the role of open-charm sector in dictating the deviation of  $\chi^{BQ}_{31}/\chi^{BQ}_{11}$  and $\chi^{BS}_{31}/\chi^{BS}_{11}$ from the HRG expectation.
 \begin{figure}[!h]
	\includegraphics[width=0.32\textwidth]{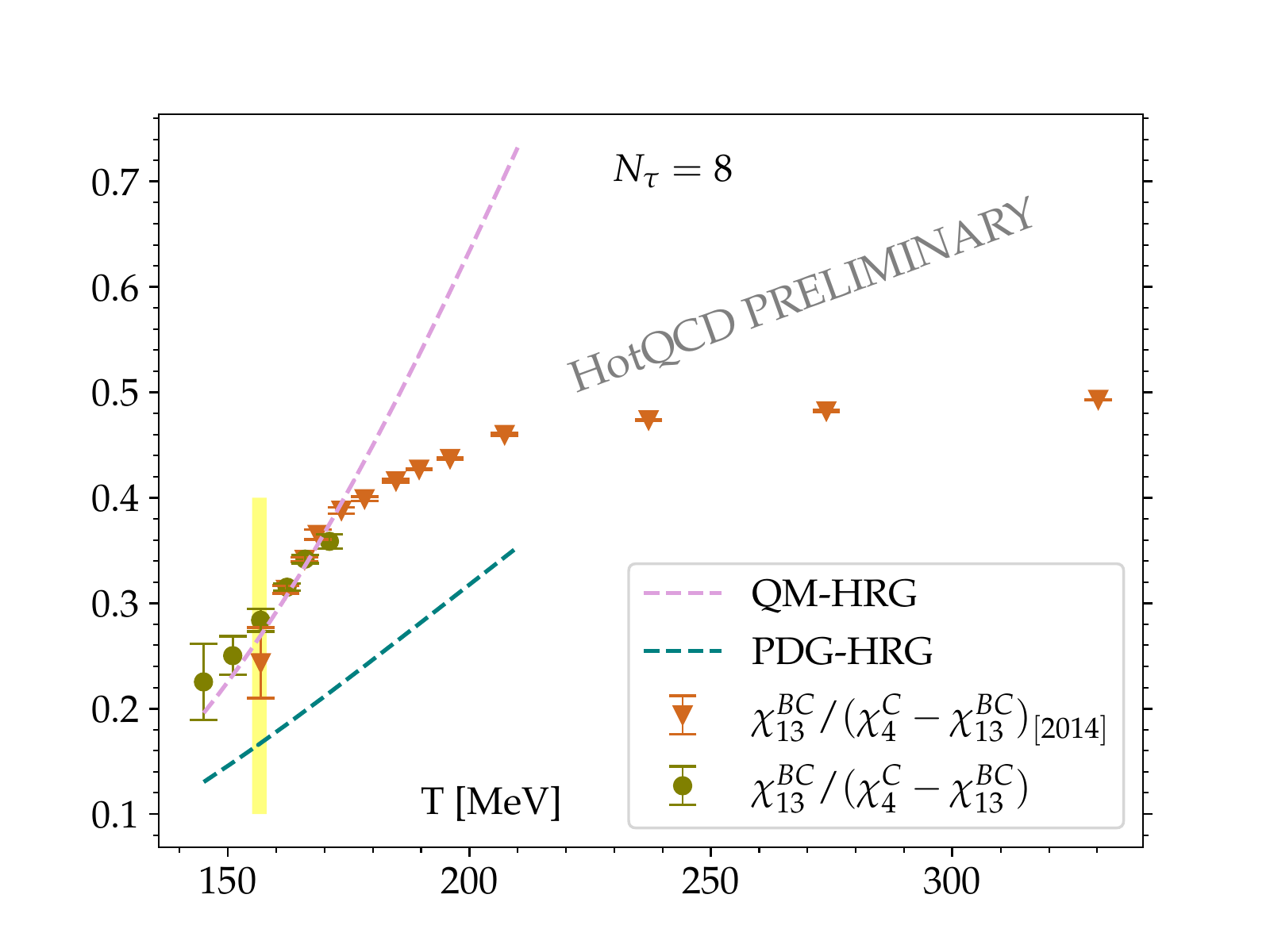}%
    \hfill
	\hspace{0.01\textwidth}%
	\includegraphics[width=0.32\textwidth]{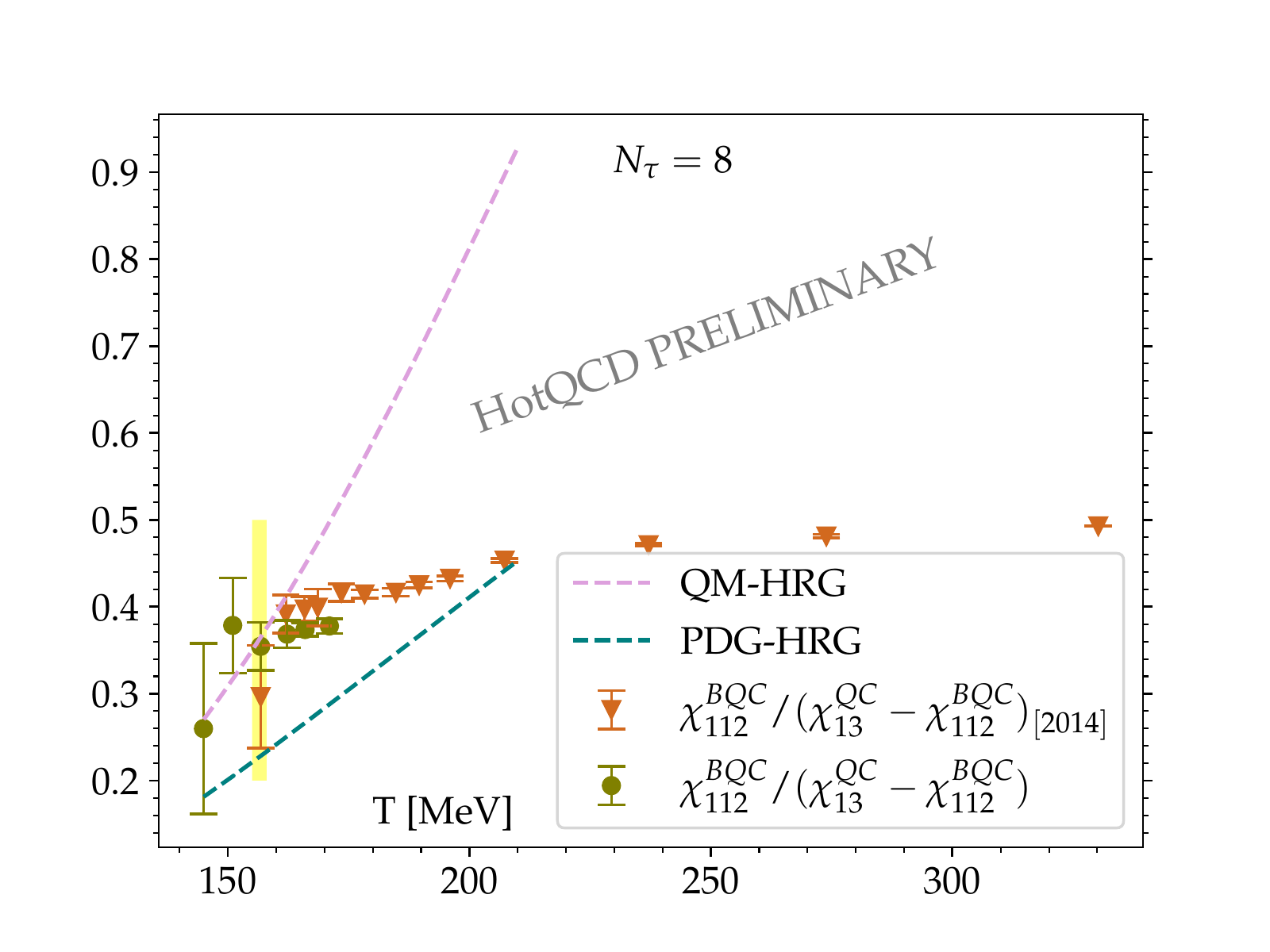}
\hspace{0.01\textwidth}
	\includegraphics[width=0.32\textwidth]{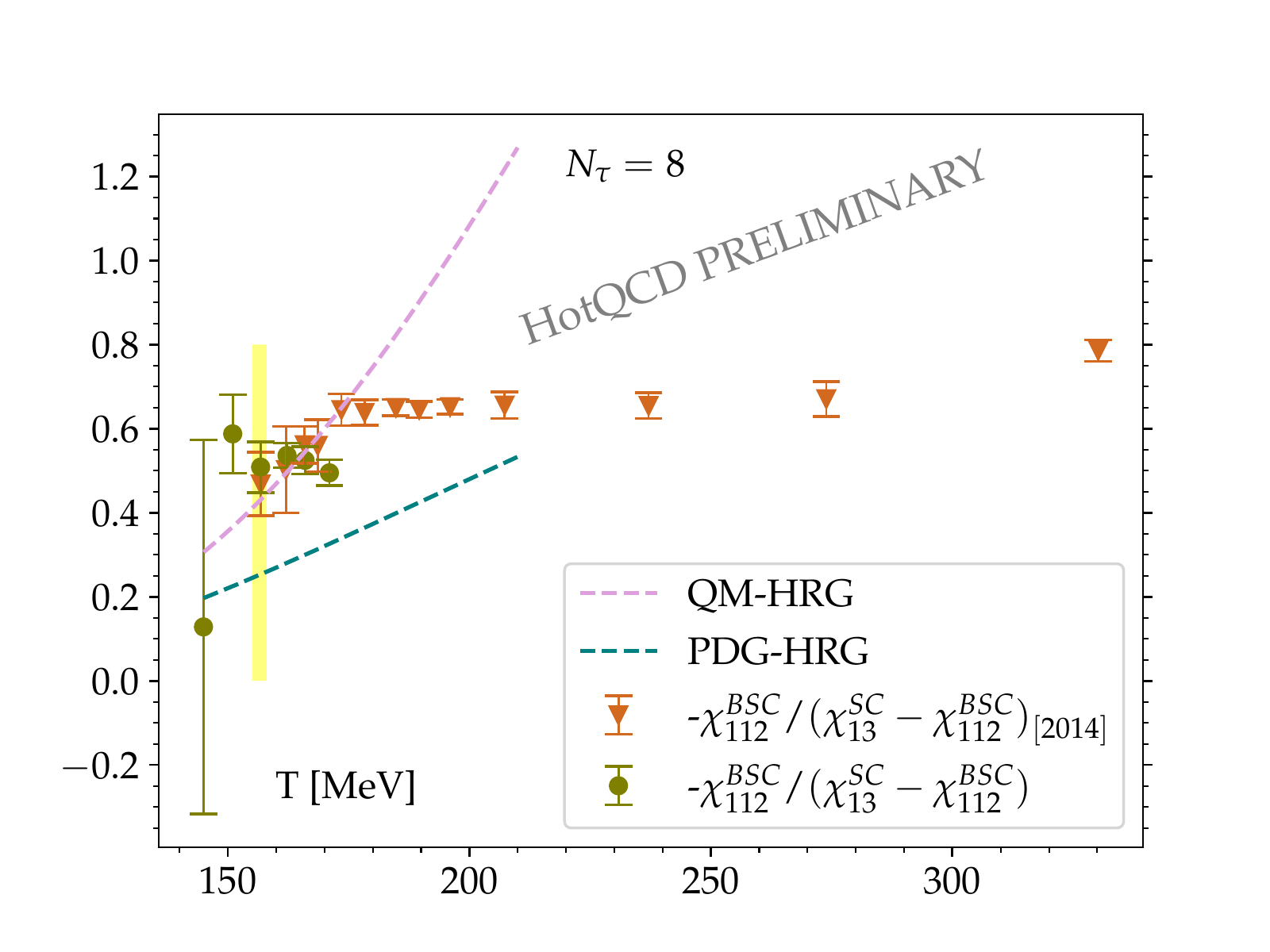}%
	\caption{Plots depicting ratios of baryonic to mesonic contribution to the partial charm pressure in charm sector [left], electrically-charged charm sector [middle], strange-charm sector [right].}
	\label{fig:BM-ratios}
\end{figure}

Quantities plotted in Fig. \ref{fig:BC-ratios} are independent of the details of the hadron mass spectrum, whereas ratios such as, $\chi^{BC}_{13}/(\chi^{C}_{4}-\chi^{BC}_{13})$ -- which effectively translates to the ratio of the contributions to the partial charm pressure from the charmed-baryonic and the charmed-mesonic sectors -- are sensitive to the details of the hadron mass spectrum. In Fig. \ref{fig:BM-ratios} [left], the dashed-teal curve representing the HRG calculation based on the states tabulated in the PDG record clearly misses the numerically calculated lattice results for $\chi^{BC}_{13}/(\chi^{C}_{4}-\chi^{BC}_{13})$, whereas upon including the states predicted via Quark-Model calculations \cite{Ebert:2009ua}, \cite{Ebert:2011kk} -- represented by the dashed-magenta curve --  the lattice results upto $T=175$ MeV can be very well described by the HRG calculation, beyond that the lattice results start to approach their non-interacting quark gas limit. Similar scenario persists for the electrically-charged charm sector tapped into using  $\chi^{BQC}_{112}/(\chi^{QC}_{13}-\chi^{BQC}_{112})$ in Fig. \ref{fig:BM-ratios} [middle], where deviation from the QM-HRG starts at $T_{pc}$. The errors of the strange-charm sector projected out using $-\chi^{BSC}_{112}/(\chi^{SC}_{13}-\chi^{BSC}_{112})$ in Fig. \ref{fig:BM-ratios} [right] are not very much under control yet for the updated data and for the 2014 HotQCD data, due to lower statistics might be getting underestimated, therefore, at present it is not possible to make a precise statement about the deviation of the lattice results from the QM-HRG calculations. Nonetheless, depiction of  the PDG-HRG curve missing the lattice results is quite vivid. In conclusion, all  sub-figures of Fig. \ref{fig:BM-ratios} reiterate the possibility of the existence of not-yet-discovered baryonic states in the open-charm sector as was argued in Reference \cite{BAZAVOV2014210}.

\acknowledgments
This work was supported by The Deutsche Forschungsgemeinschaft (DFG, German Research Foundation) - Project
numbers 315477589-TRR 211.
The authors gratefully acknowledge the computing time provided to them on the high-performance computer Noctua 2 at the NHR Center PC2 under the project name: hpc-prf-cfpd. These are funded by the Federal Ministry of Education and Research and the state governments participating on the basis of the resolutions of the GWK for the national high-performance computing at universities (www.nhr-verein.de/unsere-partner).
\bibliographystyle{unsrt}
\bibliography{refs}

\end{document}